\documentclass[pre, preprint,preprintnumbers,amsmath,amssymbl]{revtex4}

\usepackage{graphicx}
\usepackage{dcolumn}
\usepackage{bm}

\begin{document}

\title{Controlling the spectrum of x-rays generated in a laser-plasma accelerator by tailoring the laser wavefront}

\author{S.~P.~D.~Mangles$^1$}
\author{G.~Genoud$^2$}
\author{S.~Kneip$^1$}
\author{M.~Burza$^2$}
\author{K.~Cassou$^3$}
\author{B.~Cros$^3$}
\author{N.~P.~Dover$^1$}
\author{C.~Kamperidis$^2$}
\author{Z.~Najmudin$^1$}
\author{A.~Persson$^2$}
\author{J.~Schreiber$^1$}
\author{F.~Wojda$^3$}
\author{C.-G.~Wahlstr\"om$^2$}

\affiliation{$^1$Blackett Laboratory, Imperial College London, SW7 2BZ, UK}
\affiliation{$^2$Department of Physics, Lund University, P.O. Box 118, S-22100 Lund, Sweden}
\affiliation{$^3$Laboratoire de Physique des Gaz et des Plasmas, Centre National de la Recherche Scientifique, Universit\'e Paris XI, 91405 Orsay, France}

\begin{abstract}
By tailoring the wavefront of the laser pulse used in a laser-wakefield accelerator, we show that the properties of the x-rays  produced  due to the electron beam's betatron oscillations in the plasma can be controlled. 
By creating a wavefront  with coma, we find that the critical energy of the synchrotron-like x-ray spectrum can be significantly increased.
The coma does not substantially change the energy of the electron beam, but does increase its divergence and produces an energy-dependent exit angle, indicating that changes in the x-ray spectrum are due to an increase in the electron beamÕs oscillation amplitude within the wakefield.
\end{abstract} 

\maketitle

The use of intense laser pulses to excite plasma waves with a relativistic phase velocity is a possible route to the development of compact particle accelerators.   
Using this laser wakefield acceleration technique, experiments have produced quasi-monoenergetic 0.1 to 1 GeV electron beams in distances on the order of 1~cm, \cite{Leemans:NatPhys2006, Karsch:NJP2007, Kneip:PRL2009}.
Such compact particle sources have a clear potential as a source of x-rays.  
The plasma waves produced in current generation experiments not only have a strong accelerating field but also have strong focusing fields.  
These focusing fields can cause electrons within the wakefield to oscillate transverse to their acceleration direction, in `betatron orbits'.
This motion generates x-ray radiation which can have properties, in particular peak brightness, similar to those achievable with conventional `3rd generation' light sources \cite{Rousse:EPJD2007}. This is, in part, due to the ultra-short-pulse nature of these x-ray sources, which are thought to be at least as short as the laser pulse involved in the interaction, \cite{Phuoc:POP2007}, i.e. on the order of tens of femtoseconds. Such x-ray sources could be used for a wide range of studies into the structure of matter.
The ultra-short duration of the x-ray pulse and the possible femtosecond synchronization with other photon and particle sources driven by the same laser offer significant benefits.

Studies to date have concentrated on characterizing the properties of this x-ray source in terms of the spectrum, angular distribution, source size and its ultra-short nature \cite{Rousse:PRL2004, Rousse:EPJD2007, Kneip:PRL2008, Phuoc:POP2007}. 
In this letter we demonstrate an ability to control the spectral properties of the betatron x-ray source by controlling the laser wavefront.  

In Ref. \cite{Glinec:EPL2008} a correlation between the energy of the electron beam and the angle at which it exited the plasma was attributed to betatron oscillations caused by off axis injection of the electrons.
It was hypothesised that  this was caused by an aberration in the focal spot.  
However in that study no direct control of the excitation of the betratron motion was attempted, nor was the effect on x-ray production measured.
In this work we use a deformable mirror to tailor the laser wavefront and show that the x-ray spectrum can be changed significantly, in particular we show that a coma wavefront produces more high-energy photons than a `flat' wavefront.  
This change in the photon spectrum is due to an increase in  the strength of the plasma wiggler,  a direct result of off-axis injection.

In the blow-out regime relativistic electrons with energy $\gamma m_e c^2$ undergo transverse (or betatron) oscillations at the betatron frequency $\omega_\beta = \omega_p/\sqrt{2\gamma}$, with a wavenumber $k_\beta = \omega_\beta/c$ (where $ \omega_p = \sqrt{n_e e^2/ m_e \epsilon_0}$ is the plasma frequency of a plasma of electron density $n_e$).
The wiggler (or betatron) strength parameter for an electron oscillating with an amplitude $r_\beta$ is $K= \gamma k_\beta r_\beta$.
For sufficiently large oscillations ($K > 1$) the radiation is broadband and well characterized by a synchrotron-like spectrum \cite{Esarey:PRE2002, Rousse:EPJD2007}, i.e. close to the axis ($\theta = 0$) the spectrum takes the form of  
$d^2I/(dEd\Omega)_{\theta = 0} \propto \xi^2 \mathcal{K}^2_{2/3}(\xi/2)$, where $\mathcal{K}_{2/3}(x)$ is a modified bessel function of order $2/3$ and $\xi = E/E_c$.  
The shape of this spectrum is characterized by a single parameter, the critical energy $E_c$.  
The on-axis spectrum is broadband and peaked close to $E_c$, with approximately half the energy radiated below $E_c$
(note the present definition of $E_c$ is different to that used in refs \cite{Esarey:PRE2002, Kneip:PRL2008} but consistent with refs  \cite{Rousse:PRL2004, Rousse:EPJD2007, Jackson_3rd}).  
For fixed $\gamma$ and $n_e$,  the critical energy is directly proportional to the oscillation amplitude.  
$E_c = \frac{3}{4}\hbar  \gamma^2 \omega_p^2 r_\beta /  c$. 
Thus increasing the oscillation amplitude of the electrons within the wakefield is expected to have a significant effect on the x-ray photon spectrum.

The experiment was performed using the multi-TW laser at the Lund Laser Centre, which provided $0.6$~J energy pulses on target with a \textsc{fwhm} duration of $45~\pm~5$~fs at a central wavelength of $800$ nm. 
The laser was focused onto the edge of a 2 mm supersonic helium gas jet using an $f/9$ off-axis parabolic mirror. 
The plasma density was held constant at $1.5 \times  10^{19}$~cm$^{-3}$, at which the laser pulse length 
is approximately 1.5 times the  wavelength of the relativistic plasma wave $\lambda_p = 2\pi c/\omega_p = 9$ $\mu$m.

The laser wavefront was measured using a wavefront sensor (Phasics SID4).  
A 32 actuator deformable mirror (Night N Adaptive Optics DM2-65-32), placed before the focusing optic was used to correct for wavefront errors present in the laser system,  defining our 'flat' wavefront.
 The deformable mirror could be adjusted to tailor the wavefront.
The wavefront discussed here is coma, defined on a unit circle by the Zernike polynomial   
$Z(r, \theta) =  \sqrt{8}C(3r^3 - 2r)  \cos\theta$ for horizontal coma.
(where $C$ is the r.m.s. amplitude of the coma in units of the laser wavelength $\lambda$).

The intensity distribution of the focal spot was recorded on a 12 bit CCD camera using a microscope objective.  
The `flat' wavefront setting produced a focal spot with a $1/e^2$ intensity radius, or waist, of $w_0 = 12 \pm 1$~$\mu$m. 42 \% of the pulse energy was within the \textsc{fwhm}.
 This corresponds to a peak intensity of $4.0 \times 10^{18}$~Wcm$^{-2}$ or a normalised vector potential of $a_0 \simeq 1.4$, where $a_0  = eA_0/(m_e c) \simeq e E_0/(m_e \omega_0 c)$ ($A_0$ and $E_0$ are the amplitude of the vector potential and electric field of the laser and $\omega_0$ is the laser frequency).
For $C =  0.175 \lambda$ horizontal coma  the peak intensity reduced to 
$2.1 \times 10^{18}$~Wcm$^{-2}$, or a normalised vector potential of $a_0 \simeq 1$.  The energy contained within the \textsc{fwhm} reduced to 25 \%.
Horizontal coma elongates the focal spot in the horizontal ($x$) direction, but it does not do so symmetrically. 
For $C = 0.175~\lambda$ and  $x < 0$ the waist  is close to the `flat' wavefront case with 
$w_{x < 0} = 13 \pm 1$~$\mu$m. 
However, for $x > 0$ the waist is significantly increased to 
$w_{x >0} =  18 \pm 1$~$\mu$m.
Horizontal coma leaves the vertical ($y$) beam waist unaffected.

To diagnose the effect of coma on the  electron beam profile a scintillator screen (Kodak Lanex regular) was  placed in the beam path and imaged onto a 12 bit CCD camera. 
A permanent magnet ($B = 0.7$~T, length 100 mm)
could be moved into position between the gas jet and the scintillator screen to sweep the electrons away from the laser axis.  
The magnetic field dispersed the electrons in the vertical direction, the vertical ($y$) position of the electron beam on the scintillator screen is then a function of the beam energy ($E$) so that, taking into account the non-linear dispersion $dy/dE$ and the response of the lanex to high energy electrons \cite{Glinec:RSI2006} the electron energy spectrum $dN/dE$ can be calculated.

X-rays generated by betatron oscillations in the wake were recorded by an x-ray sensitive 16 bit CCD camera (Andor 434-BN).
The CCD chip had $1024 \times 1024$ pixels and was placed on the laser axis.  
The chip collection angle corresponded to  $20~$mrad $\times~20$~mrad.
The x-rays were only recorded when the magnet was in position to prevent the electron beam striking the CCD chip.
An array of filters (Al, Zn, Ni, Fe) were placed directly in front of the CCD chip.   
Using the known transmission of x-rays \cite{NIST} through the filters, and the CCD sensitivity, we find the critical energy $E_c$ which best describes the  x-ray photon spectrum using a least-squares method \cite{Kneip:PRL2008}.

\begin{figure}
\begin{center}
\includegraphics[width=8.5cm]{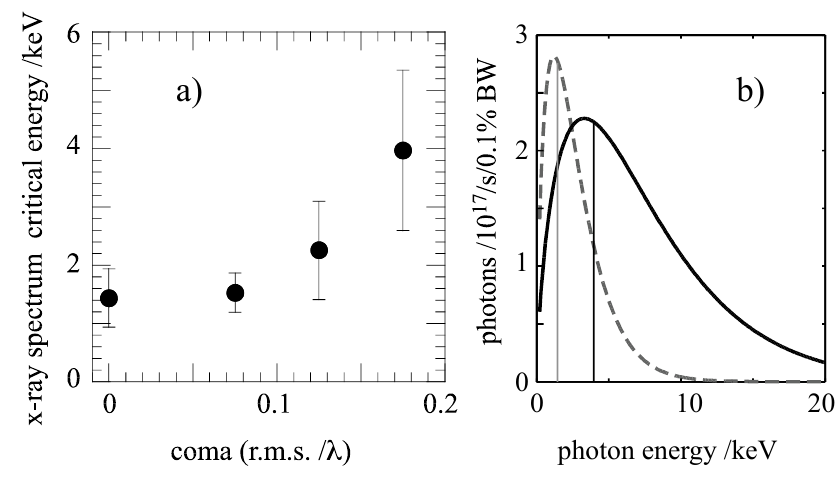}
\caption{a) Variation of the observed x-ray spectrum critical photon energy, $E_c$ with the amplitude of coma.  For larger coma no x-rays or were observed.
b) best-fit synchrotron spectra for a `flat'  wavefront (dashed line) and 0.175 $\lambda$ amplitude coma (solid line).  The curves take into account the expected change in beam divergence and assume the duration of the x-rays is that of the laser pulse. The vertical lines indicate the position of $E_c$.}
\label{Ec-coma}
\end{center}
\end{figure}

The variation of $E_c$ with coma amplitude is shown in Fig. \ref{Ec-coma}a).  
For the `flat' wavefront we observed $E_c = 1.5  \pm 0.5$~keV.  
As the coma amplitude increases, a clear shift in the x-ray spectrum towards higher photon energies is observed, reaching   $E_c = 4.0 \pm 1.5 $~keV for 0.175 $\lambda$ coma.  
The electron spectra shows a constant mean electron energy of $\langle\gamma \rangle =  144 \pm 7 $ for all the coma settings, implying that  the change in $E_c$ is due to a change in the oscillation amplitude, $r_\beta$.  
We calculate that the oscillation amplitude increases from  $r_\beta  = 1.0 \pm 0.4 $~$\mu$m to $r_\beta = 3 \pm  1$~$\mu$m corresponding to an increase in the wiggler strength parameter  from $K = 5 \pm 2 $ to $K =  17 \pm 6$. 
The largest oscillation amplitudes are slightly less than the radius of the wakefield (approximately $\lambda_p/2 = 4.5$~$\mu$m) and indicate that almost the whole width of the plasma channel is being used for radiation generation. 
Oscillations larger than this would not be supported and may indicate why, when larger amplitude coma wavefronts were used, no electrons or x-rays were observed.

Integrating the signal recorded on the CCD and taking into account the expected photon beam divergence,  $\theta \approx K/\gamma$, we estimate that as the coma is increased the number of photons remains approximately constant at  $(3 \pm 1) \times 10^{7}$ photons per shot. The energy in the x-ray beam therefore increases, as expected for an increase in $K$.
Two x-ray spectra are shown in Fig. \ref{Ec-coma}b).
While the peak intensity is reduced for the case of 0.175~$\lambda$ coma, there is significantly more intensity at the higher photon energies.

\begin{figure}
\begin{center}
\includegraphics[width=8.5cm]{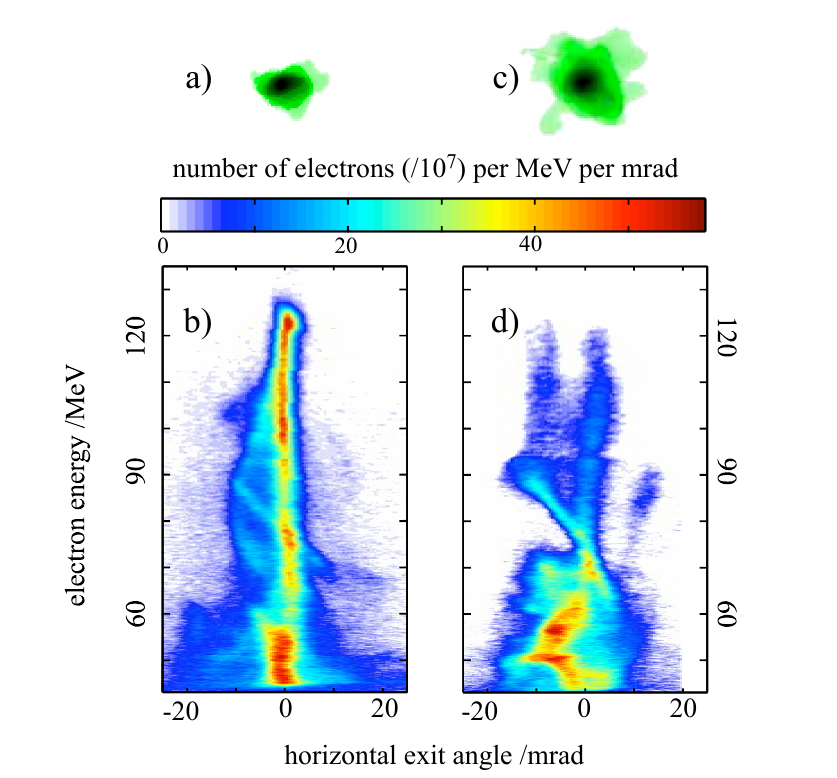}
\caption{(color online) Effect of coma on the electron beam.  a, b) `flat' wavefront, c, d) coma =  0.175 $\lambda$, a \& c) show the electron beam profile (average from several shots)
b \& d) show the electron energy spectrum (from a single shot). The vertical axis is a linear energy scale and the horizontal scale represents the exit angle in the non-dispersion plane.}
\label{especs}
\end{center}
\end{figure}

The source of the change in the x-ray spectrum  can be elucidated  by examining the effect of coma on the electron beam.
Figure \ref{especs} shows the variation of the electron beam profile and energy spectrum with the amplitude of coma. 
The electron beam profile images (each representing an average of five shots) show that the electron beam divergence increases with the amplitude of coma from $\simeq 10$~mrad (\textsc{fwhm}) for the `flat' wavefront to $\simeq 20$~mrad for 0.175 $\lambda$ coma.
The images of the electron energy spectrum (each from a single shot) show the electron energy spectrum $dN/dE$ as a function of the horizontal angle at which the beam exits the plasma
For shots with the `flat' wavefront and  $n_e = 1.5 \times 10^{19}$~cm$^{-3}$ the electron spectrum is broadband, extending in energy to $\simeq  120$~MeV. 
For the `flat' wavefront the electrons have an approximately constant exit angle, indicating a small betatron amplitude.  
With coma we observe an energy dependent exit angle, indicating a large betatron oscillation amplitude.  
This occurs because, for broad energy-spread beams, an electron's energy can be mapped to its phase within the plasma wave.  
Different energy electrons will therefore be at  a different phase of their betatron orbit as they exit the plasma, resulting in an oscillatory dependence of the exit angle with energy,  as observed.

The electron beam diagnostics show that, by tailoring the wavefront to create an asymmetric focal spot, the wake dynamics are sufficiently perturbed so as to increase the amplitude of the betatron oscillations. 
This is likely due to the promotion of off-axis injection due to the generation of an asymmetric  wakefield by the asymmetric focal spot.
Images of self-scattered radiation   show that a long wavelength hosing of the channel sometimes occured;
the likelihood of observing this hosing increased with coma.   
However, the hosing wavelength was significantly longer than the plasma wavelength so would not produce an energy dependent exit angle, and no correlation between the amplitude of the hosing and x-ray spectrum was observed.

By tailoring the laser wavefront to produce an asymmetric  focal spot we have demonstrated an ability to change the energy spectrum of the x-ray photons from a laser-plasma wiggler; increasing the number of high-energy ($E \gtrsim 5$~keV) photons without the need to increase the laser power or electron beam energy.  This offers an alternative route to higher energy laser-based x-ray sources without the significant cost of PW laser facilities.  
On some shots we observe that the effect of the coma on the electron beam divergence is predominantly in the  plane of the laser asymmetry. 
If this effect can be controlled then it also offers a mechanism for  producing polarized x-rays with a laser plasma wiggler.

We acknowledge support from The Royal Society, the Marie Curie Early Stage Training Site \textsc{maxlas} (\textsc{mest-ct}-205-02936), the Swedish Research Council and the  Euro\textsc{leap} network.

\bibliography{coma-xrays5}

\end{document}